\documentclass{moriond}
\pdfoutput=1
\usepackage{amsmath,amssymb,amsfonts,graphicx,rotating,hyperref,slashed,xcolor,cite,expdlist}
\makeatletter
\usepackage{lmodern,textcomp}
\usepackage{soul}
\usepackage{blkarray}
\usepackage{multirow,multicol}
\usepackage{dsfont}
\usepackage{footnote}
\definecolor{rosso}{cmyk}{0,1,1,0.4}
\definecolor{rossos}{cmyk}{0,1,1,0.55}
\definecolor{rossoc}{cmyk}{0,1,1,0.2}
\definecolor{blu}{cmyk}{1,1,0,0.3}
\definecolor{blus}{cmyk}{1,1,0,0.6}
\definecolor{bluc}{cmyk}{1,1,0,0.1}
\definecolor{verde}{cmyk}{0.92,0,0.59,0.25}
\definecolor{verdec}{cmyk}{0.92,0,0.59,0.15}
\definecolor{verdes}{cmyk}{0.92,0,0.59,0.4}
\hypersetup{colorlinks,bookmarksopen,bookmarksnumbered,
linkcolor=blus,pdfstartview=FitH,urlcolor=rossos,citecolor=verde}
\allowdisplaybreaks

\def\be{\begin{equation}}
\def\ee{\end{equation}}
\def\bea{\begin{eqnarray}}
\def\eea{\end{eqnarray}}

\newcommand{\Photo}{\includegraphics[height=30mm]{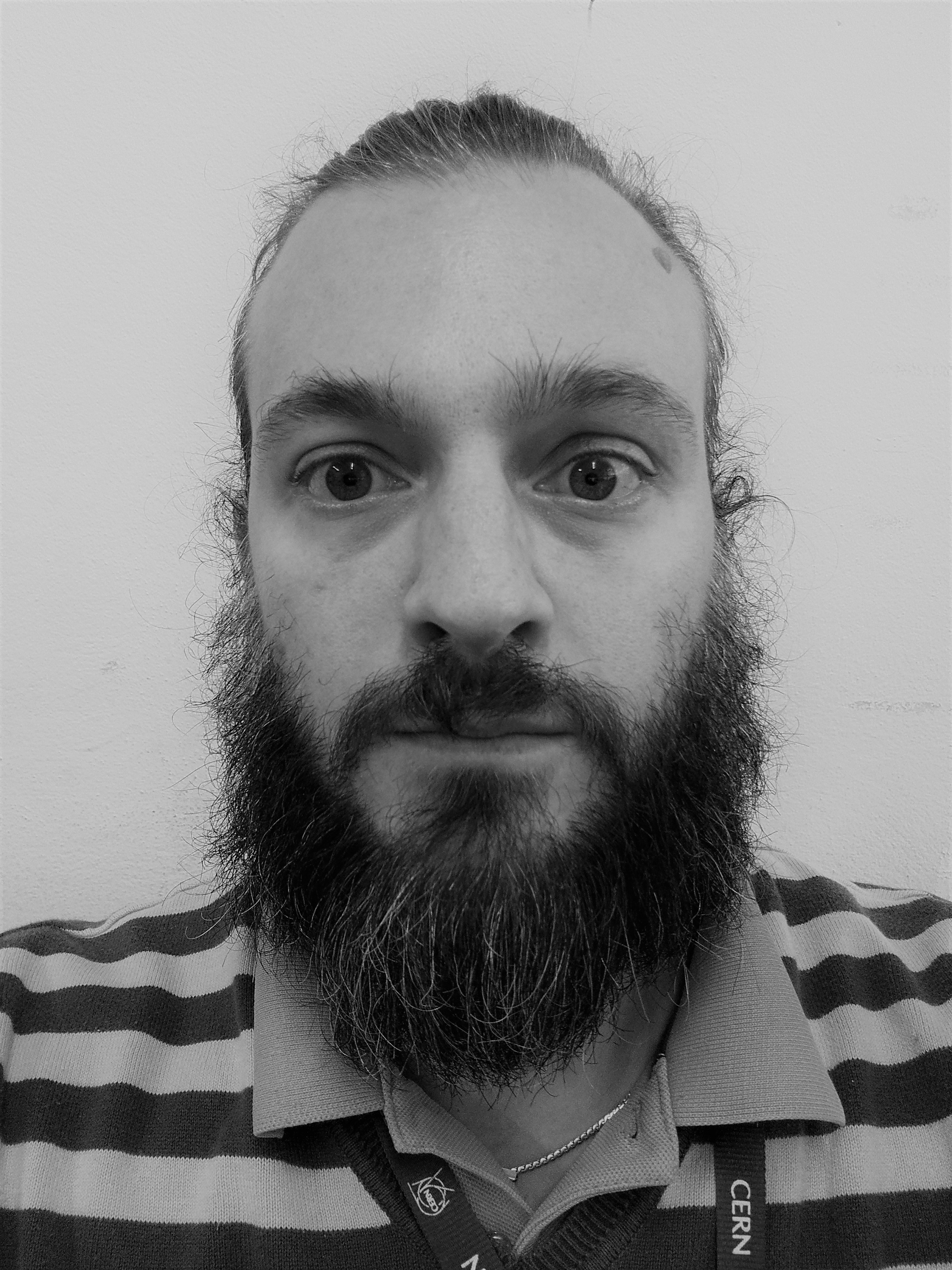}}

\begin{document}
\title{CLOCKWORK/LINEAR DILATON: STRUCTURE AND PHENOMENOLOGY}

\author{RICCARDO TORRE}

\address{CERN, Theory Division, Geneva, Switzerland {\rm and} INFN Sezione di Genova, Italy\vspace{2mm}}

\maketitle\abstracts{
I briefly discuss the main phenomenological features and constraints of the Clockwork/Linear Dilaton (CW/LD) 5D geometry.\\
This contribution is based on the work of ref.~\cite{Giudice:2017fmj}, to which the reader is referred for an extensive discussion of the subject and the full list of relevant references. The only original result of this proceeding is adding the constraint arising from the CMS analysis \cite{CMS:2018thv} to the summary plot.}


\section{Introduction}

\noindent Why is the Fermi constant $G_{F}\approx 1.2\cdot 10^{-5} \text{GeV}^{-2}$ so much larger than the Newton constant $G_{N}\approx 6.7\cdot 10^{-39} \text{GeV}^{-2}$ is one of the biggest mysteries of Nature. This question can be considered as a qualitative formulation of the electroweak Hierarchy Problem (HP), that is, why are weak interactions so much stronger than gravitational interactions? Indeed there are three crucial scales known in particle physics,\footnote{There is indication for other scales related to neutrino masses, gauge coupling unification, strong CP problem, and so on, but there is no direct evidence of any of them at the time of this writing.} one of which, the QCD scale, is dynamically generated by strong interactions. The other two, the weak scale, related to the Higgs boson mass $m_{H}$ by the coupling constant $\sqrt{\lambda}$ through $1/G_{F}=\sqrt{2} v^{2}=m_{H}^{2}/(\sqrt{2} \lambda)\approx (300 \text{GeV})^{2}$ and the Planck scale $1/G_{N} = M_{P}^{2}\approx (10^{19}\text{GeV})^{2}$, are unexpectedly far apart. In the absence of any symmetry and/or dynamical mechanism responsible for the generation of these scales, these two parameters both break scale invariance, and one would expect them to be roughly of the same order. This is what Naive Dimensional Analysis (NDA) would suggest. For this reason many physicists believe that some symmetry and/or dynamical mechanism should be responsible for this very large scale separation. This could be Supersymmetry (SUSY), dimensional transmutation, localization in extra-dimensional space-time, or any other mechanism that completely forbids, or makes irrelevant in the ultraviolet (UV), the Higgs mass operator.

Extra-dimensions offer a useful and clean perspective on the HP \cite{Rattazzi:2003ea}. The issue of the weakness of gravity, or the stability of the Higgs potential under radiative corrections, is turned into an issue of stability of a higher dimensional space-time geometry. In its simplest realization the Standard Model (SM) fields are localized on a 4-dimensional hypersurface of a $d>4$ higher-dimensional space-time and the fundamental $d$-dimensional Planck scale is assumed to be of the order of the electroweak scale $M_{P}^{(d)}\sim 1/G_{F}$, so that no hierarchy is present between the two fundamental scales. The observed weakness of gravity, corresponding to a much larger value of the 4D Planck scale $M_{P}\gg M_{P}^{(d)}$, could then be explained by ``dilution'' of the gravitational force in a large extra-dimensional volume. This large volume can arise in two different ways: it could correspond to very large extra-dimensions with fixed 4D volume along the extra coordinates, that is flat and large extra-dimensions, or to small extra-dimensions where the 4D volume grows along the extra-coordinates, referred to as warped extra-dimensions. 

Both the mechanisms illustrated above give a simple solution to the weakness of gravity, being a potential solution to the HP. However, one needs to explain the mechanism that generates such higher dimensional geometries and make sure that it does not require a large fine-tuning.

The simplest example is a single, flat, and large extra-dimension \cite{ArkaniHamed:1998rs,Antoniadis:1998ig,ArkaniHamed:1998nn}. It is well known from the flatness problem of our universe that explaining a very flat geometry requires the tuning of a very small Cosmological Constant (CC). Indeed, if we assume that the fundamental Planck scale of an extra-dimensional flat space is of the order of the electroweak scale, then we naturally expect a CC of the order of $m_{H}^{2}/G_{F}$, which in turn would imply a geometry that is either de Sitter (dS) or Anti-de Sitter (AdS). Getting a very flat extra-dimension would correspond to finely tuning the bulk CC. This problem has nothing to do with the existing problem of tuning the 4D CC to get 4D Poincar\'e invariance and is nothing but the translation of the electroweak HP into a problem of stability, or naturalness, of the extra-dimensional geometry. Therefore, to represent a solution to the HP, an extra-dimensional model needs a mechanism to stabilize the extra-dimension, or in other words, to realize the required geometry without finely tuning additional parameters. It turns out that it is generally harder to stabilize a flat (Poincar\'e) rather than a warped (e.g.~dS or AdS) extra-dimension. The flatness is strictly related to the vanishing of the bulk CC, which is known to be possible only in the presence of SUSY (no other symmetry is known to prevent the existence of a CC and be preserved by radiative corrections). Moreover, a single, flat, large extra-dimension would need to have solar system size to explain the observed hierarchy, and would be completely ruled out by observations. For this reason one usually needs more than one flat extra-dimension and the model complicates further. 

Naturally realizing warped extra-dimensions, of the Randall-Sundrum (RS) type ~\cite{Randall:1999ee}, i.e. inspired by the AdS/CFT correspondence, is usually simpler. The fine-tuning in realizing the needed AdS 5D geometry corresponds to fixing the size of the extra-dimension, that is the distance between the IR and UV branes. This distance corresponds to the vacuum expectation value of a dynamical field in the bulk, the radion. If the distance is not fixed (stabilized), the radion remains massless and mediates a long range force that leads to unacceptable modifications of the gravitational potential. Generating dynamically a potential (a mass) for the radion fixes the distance between the two branes. Being able to dynamically fix this distance at the value required by naturalness without having to fine-tune 5D parameters gives a natural solution to the HP. The so-called radius stabilization, which turns a rephrasing of the HP into an actual solution of it, can be achieved with a simple mechanism, known as Goldberger-Wise (GW) mechanism~\cite{Goldberger:1999uk}. A new scalar is added to the model with mass terms both in the bulk $m$ and on the boundaries $\mu_{i}$. A mild hierarchy between the bulk GW mass $m$ and the inverse AdS radius $1/R$, $mR\sim 1/30$ is sufficient to realize the observed electroweak-gravity hierarchy. In the dual language of the CFT the bulk mass of the GW scalar corresponds to the dimension of a nearly marginal deformation of the UV CFT. The RS geometry is only a prototype of warped extra-dimensional solutions to the HP. One can construct different geometries that share the same behavior of RS in the UV, but flow to different geometries in the IR. In this way the holographic interpretation of radius stabilization is not lost, but the physics, that means the spectrum and couplings of KK-modes, close to the IR brane where the SM fields live, can be different. 

Recently an extra-dimensional warped geometry called Linear Dilaton (LD), that is not asymptotically AdS in the UV, has been singled out from different directions as a new candidate extra-dimensional solution to the HP. On the UV side it has been shown to arise from a duality different than AdS/CFT and much less understood: LD backgrounds emerge as the 7D gravitational dual \cite{Aharony:1998ub,Giveon:1999px} to Little String Theory (LST)~\cite{Berkooz:1997cq,Seiberg:1997zk}, that is a 6D strongly-coupled non-local theory arising on a stack of NS5 branes. Explaining this duality is beyond any scope of this proceeding, it is sufficient to notice that 5D deconstructions of these 7D LD setups have been introduced inspired by this string theory construction \cite{Antoniadis:2001sw}. On the IR side instead, the same 5D geometry has been recently obtained as a particular continuum limit of a clockwork model with infinite sites \cite{Giudice:2016yja,Craig:2017cda,Giudice:2017suc}. For this reason, this geometry has been recently renamed \mbox{Clockwork/Linear Dilaton (CW/LD)}. It is essential to notice that in both cases a bulk CC, allowed by the symmetries of the model, has not been introduced. In the string theory inspired construction this is due to the assumption of SUSY in the bulk, while in the CW construction is due to the assumption of vanishing cosmological constants on the 4D sites. As it is extensively discussed in ref.~\cite{Giudice:2017fmj} the bulk cosmological constant constitutes the main obstacle in the stabilization of the extra-dimension: indeed it would make the model asymptotically AdS in the UV, and it will very quickly change its salient features back to RS-like ones. This suggests that SUSY may be a necessary ingredient in explaining the absence of a 5D CC (see ref.~\cite{Teresi:2018eai} for a different proposal that avoids SUSY). It is likely that the mechanism of radius stabilization in the CW/LD is related to the LD/LST duality, as it happens in the Randall-Sundrum theories, and that a deeper understanding of one of these two dual theories, will shed light on the other.

Going beyond the stabilization mechanism, that is certainly crucial, but is also conceptually more complicated than in the case of RS, and that still claims for a clear understanding, the CW/LD geometry has singled out also for its radically different phenomenological features. Studies of the CW/LD phenomenology can be found in refs.~\cite{Antoniadis:2011qw,Baryakhtar:2012wj,Cox:2012ee,Giudice:2017fmj}.
The scope of the next few pages is to briefly list the salient features of the phenomenology of the CW/LD model and to summarize constraints on its parameter space.

\section{The LD model}

The full action of the model is given in ref.~\cite{Giudice:2017fmj}. Einstein's equations and the dilaton equation of motion deriving from this action are solved by the metric
\be
ds^2 = e^{\frac43 k |y|}\left(\eta_{\mu\nu}dx^\mu dx^\nu + dy^2\right) ,\qquad
S(y) = 2 k |y| \, ,
\label{eq:ScalarBackground}
\ee
which defines the CW/LD geometry.
This metric implies the following relation between the fundamental scale $M_{5}$, that defines the cut-off of the theory, and that is supposed to be not much larger than the electroweak scale, and the four-dimensional reduced Planck mass, $M_P \equiv 1/\sqrt{8\pi G} \approx 2.4 \times 10^{18}$~GeV:
\be
M_P^2 = \frac{M_5^3}{k}\left(e^{2\pi kR} - 1\right) \, .
\label{MPl}
\ee
The required electroweak/gravity hierarchy is generated for
\be
kR \simeq \frac{1}{\pi}\ln\left(\frac{M_P}{M_5}\sqrt{\frac{k}{M_5}}\right) \approx 10 +\frac{1}{2\pi} \ln \left( \frac{k}{\rm TeV}\right) -
\frac{3}{2\pi} \ln \left( \frac{M_5}{10~{\rm TeV}}\right) \, .
\label{kR}
\ee
The issue of the naturalness of this choice is related to the mechanism of radius stabilization, and is extensively discussed in ref.~\cite{Giudice:2017fmj}.

\begin{figure}[t]
\begin{center}
\includegraphics[width=0.43\textwidth]{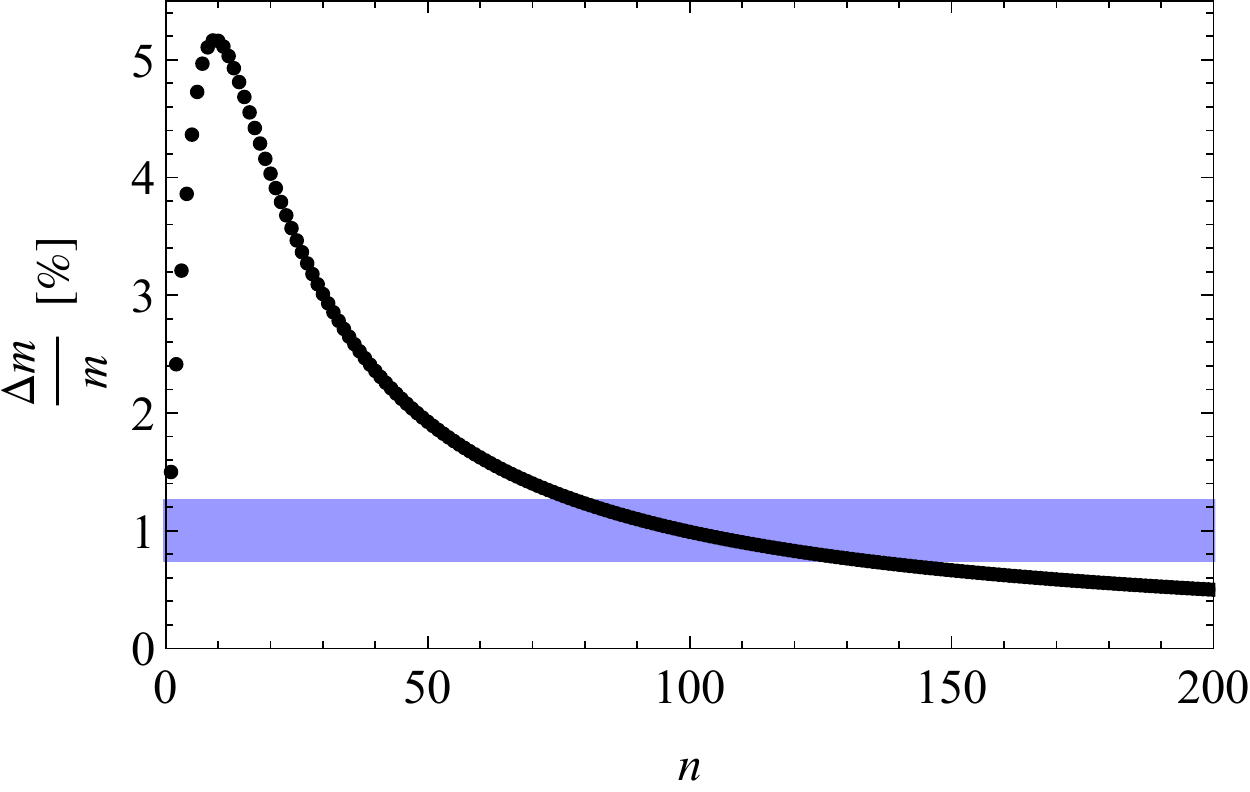}\hspace{4mm}
\includegraphics[width=0.463\textwidth]{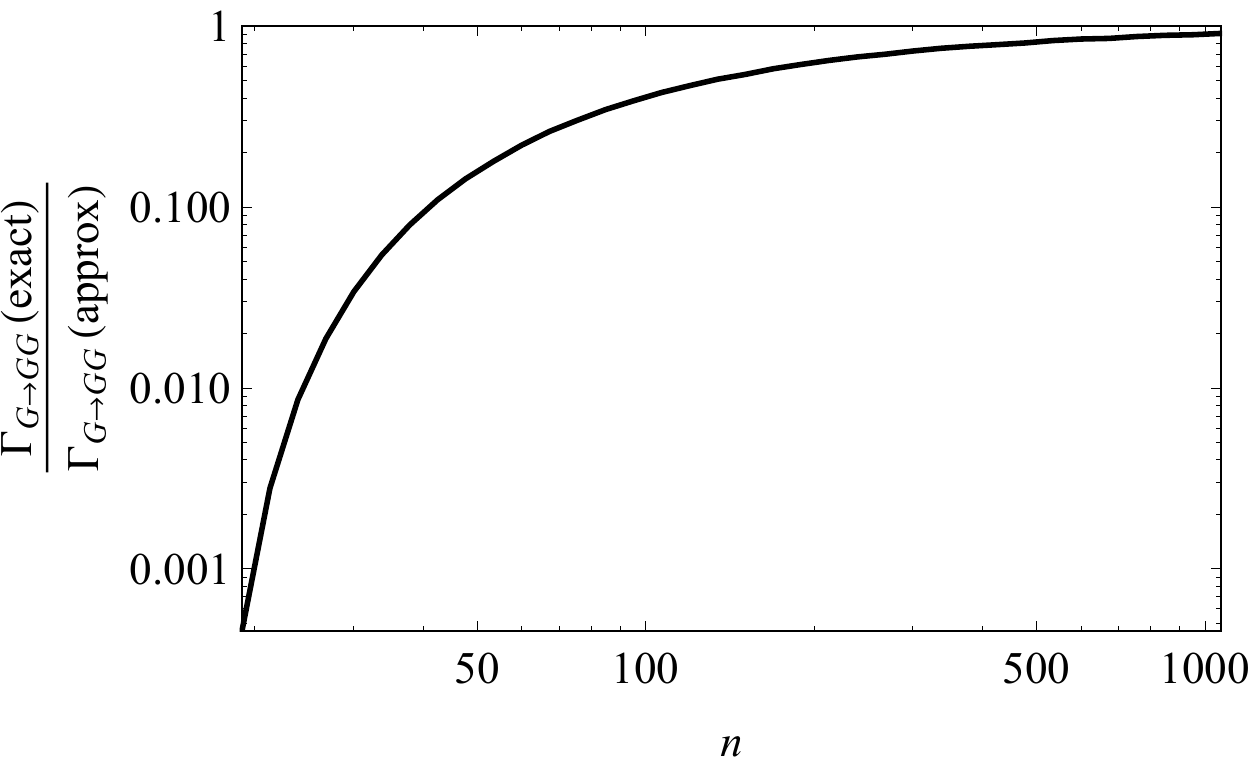}
\end{center}\vspace{-4mm}
\caption{Left panel: KK graviton mass splittings as a function of the mode number. The behavior is shown for $M_5 = 10$~TeV, $k = 200$~GeV, however the result does not strongly depend on these parameters. The blue band indicates the typical experimental resolution in the di-photon and di-electron channels ($\approx 1\%$). Right panel: exact partial width for the decay of a mode-$n$ KK graviton into pairs of lighter KK gravitons relative to the asymptotic expression given in eq.~\eqref{approx-KKG-KKG}. Here we have fixed $M_5 = 10$~TeV, $k = 10$~GeV, but the dependence on these parameters is mild.}\vspace{-3mm}
\label{dm}
\end{figure}

The spectrum of Kaluza-Klein (KK) gravitons is given by
\be
m_0 = 0\,,\qquad
m_n^2 = k^2 + \frac{n^2}{R^2}\,,\qquad
n \in \mathbb{Z}\,,
\label{KKG-masses}
\ee
so that the KK modes are separated from the zero mode, that is the massless graviton, by a mass gap of order $k$. A peculiar feature of this model is that the KK modes form a narrowly-spaced spectrum above the mass map. The left panel of figure~\ref{dm} shows the relative mass splitting as a function of the KK number $n$. This splitting drops below the di-photon and di-lepton mass resolution of around 1\% for $n\gtrsim 100$.

In the CW/LD the SM fields are localized on the IR brane. Therefore KK gravitons couple to the SM via $T^{\mu\nu}$ as
\be
{\cal L} \supset -\frac{1}{\Lambda_{G}^{(n)}}\, \tilde h^{(n)}_{\mu\nu}\,T^{\mu\nu} \,,\quad 
\Lambda_G^{(0)2} = M_P^2\,,\qquad
\Lambda_G^{(n)2} = M_5^3\pi R\left(1+\frac{k^2R^2}{n^2}\right) = M_5^3\pi R\left(1-\frac{k^2}{m_n^2}\right)^{-1} \, .
\label{KKgraviton-couplings}
\ee
so that these couplings are not suppressed by $M_P$, but by a much lower scale. 

These interactions lead, for KK gravitons much heavier than the SM particles, to the following branching ratios for the KK gravitons decays:

\vspace{2mm}
\begin{center}
\begin{tabular}{c|c|c|c|c|c|c|c}
$gg$ & $\sum_i q_i\bar q_i$ & $W^+W^-$ & $ZZ$ & $hh$ & $\gamma\gamma$ & $\sum_i\ell_i^+\ell_i^-$ & $\sum_i\nu_i\bar{\nu}_i$ \\\hline
34\% & 38\% & 9.2\% & 4.6\% & 0.35\% & 4.2\% & 6.4\% & 3.2\%
\label{BRs}
\end{tabular}\,.
\end{center}\vspace{2mm}

The detailed expressions, including phase space effects, which must be taken into account for lighter KK gravitons, can be found in ref.~\cite{Giudice:2017fmj}. The total decay rate of a mode-$n$ KK graviton into SM particles in the same limit is
\be
\Gamma_{G_n \to {\rm SM}} = \frac{283}{960\pi}\,\frac{m_n^3}{\Lambda_G^{(n)2}}
= \frac{283}{960\pi^2}\frac{m_n^3}{R M_5^3}\left( 1-\frac{k^2}{m_n^2}\right) \, .
\ee
In the absence of other decay channels, KK graviton decays can be prompt, displaced, or even stable at detector scale, especially for large values of $M_{5}\gtrsim 10$~TeV. 
As a result, it is possible that, within the same theory at a given $M_5$ and $k$, some KK gravitons decay promptly, while others lead to displaced vertices, which is a very peculiar signature of this model.

\begin{figure}
\begin{center}
\includegraphics[width=0.37\textwidth]{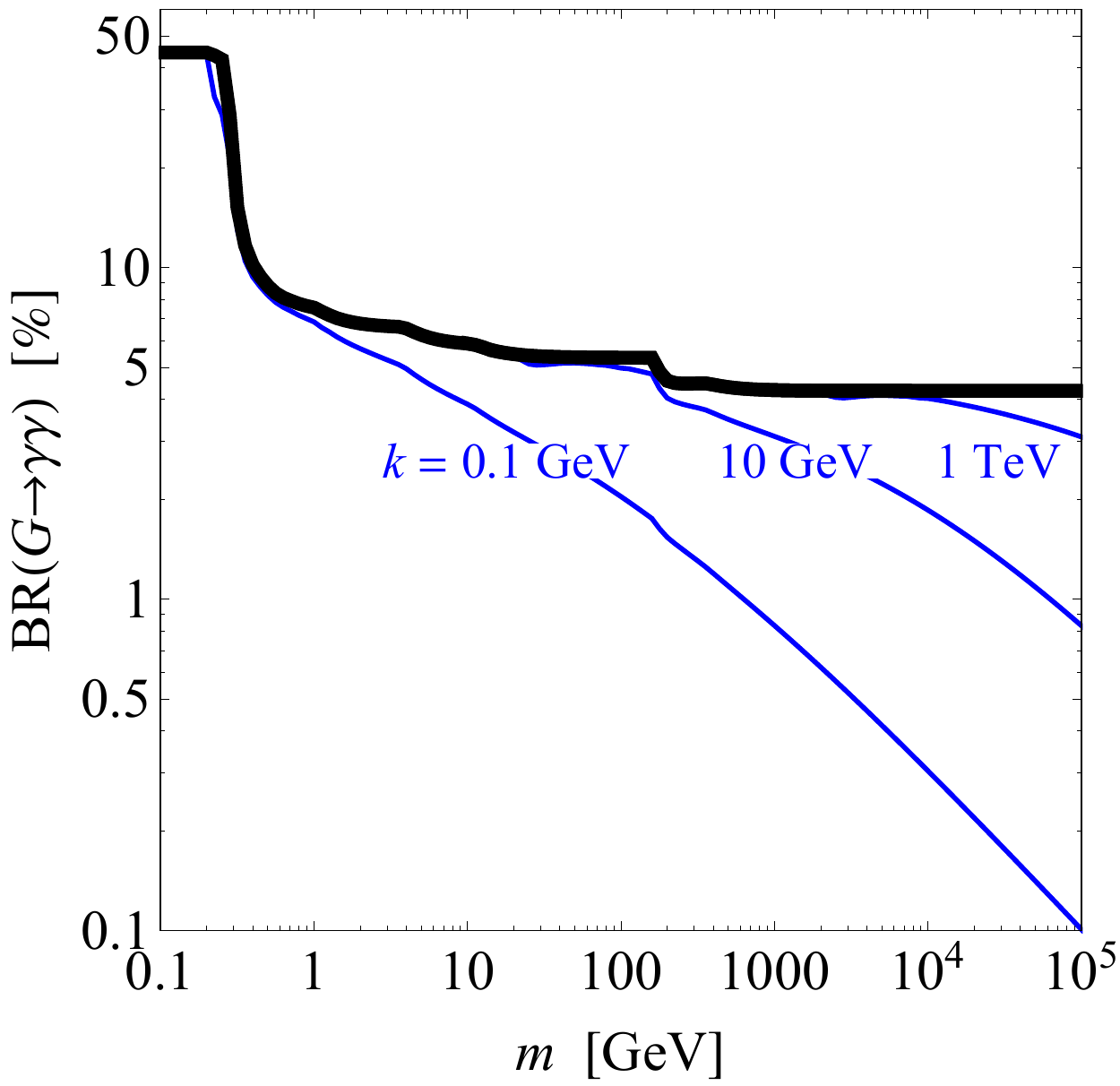}\hspace{1cm}
\includegraphics[width=0.37\textwidth]{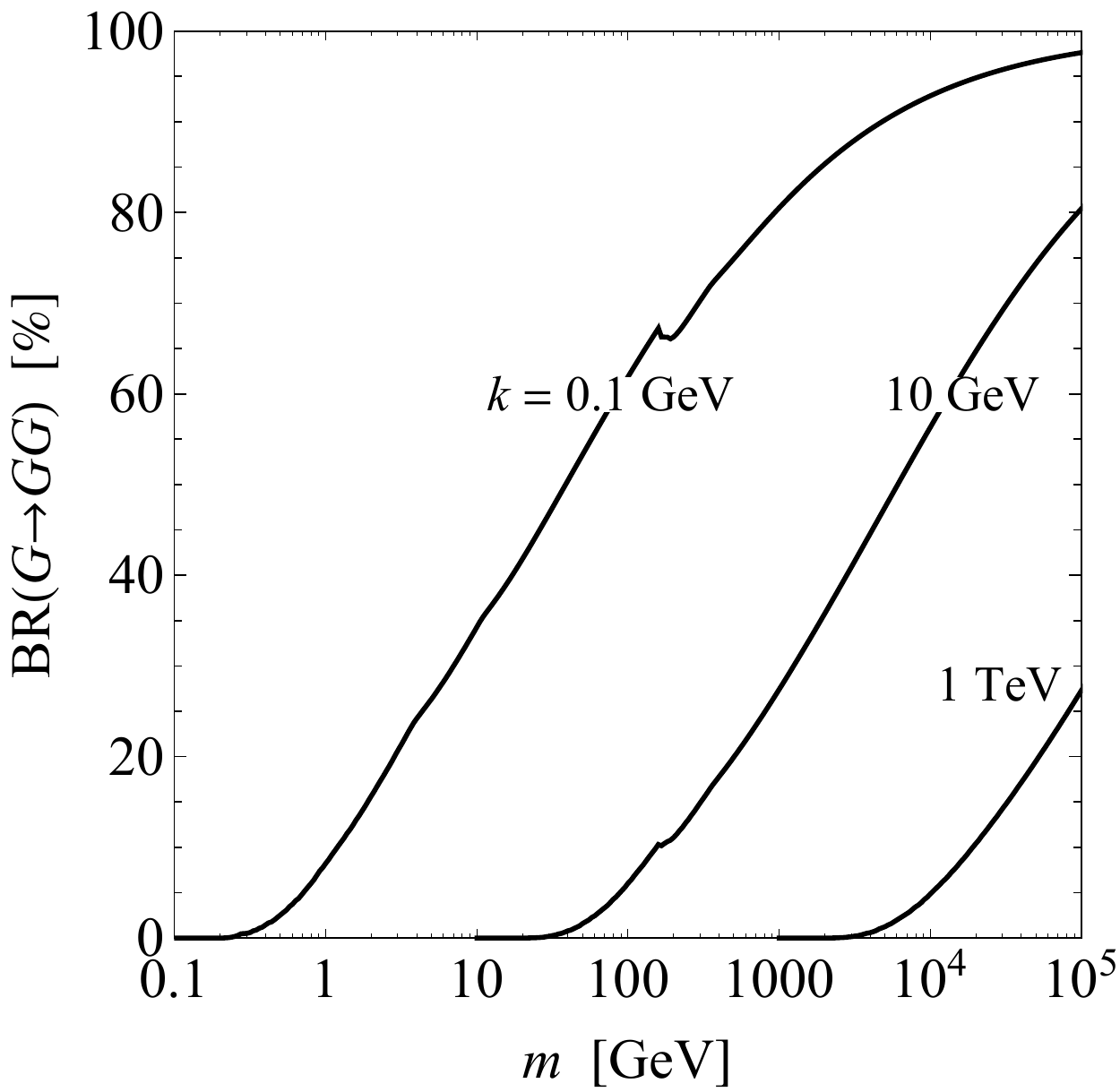}
\end{center}\vspace{-4mm}
\caption{KK graviton branching fractions to photons (left) and to lighter KK gravitons (right) for $k = 0.1$, 10 and 1000~GeV, as a function of the KK graviton mass. In the left plot, the thick black curve shows the result that would be obtained without accounting for decays to lighter KK modes.}\vspace{-3mm}
\label{BRs-w-KKG-KKG}
\end{figure}

In the CW/LD model 5D graviton self-interactions lead in 4D to KK graviton decays into pairs of lighter KK gravitons.
Details on this process can be found in ref.~\cite{Giudice:2017fmj}. Here we limit ourselves to writing down their total rate in the limit $n \gg kR \gg 1$:
\be
\Gamma_{G_n \to \sum G_\ell G_m} \simeq 
\frac{595}{3 \times 2^{14}\, \pi^2}\,
\frac{m_n^{7/2}}{k^{1/2}RM_5^3} \, .
\label{approx-KKG-KKG}
\ee
The deviation of this approximate formula from the exact result is shown in the right panel of figure \ref{dm} as a function of the KK number $n$.
This rate becomes large for large $n$ and can even become the dominant one:
\be
\frac{\Gamma_{G_n \to \sum G_\ell G_m}}{\Gamma_{G_n \to {\rm SM}}} 
\approx 4.1\times 10^{-2}\sqrt{\frac{m_n}{k}} \;.
\ee
This can have a big impact on the phenomenology: on the one hand, BRs into SM particles are decreased, as shown for instance in the case of the di-photon channel in figure~\ref{BRs-w-KKG-KKG} (left); on the other hand, decays to lighter KK gravitons can give important high multiplicity signatures, since the branching fraction for such decays can be large, as shown in figure~\ref{BRs-w-KKG-KKG} (right).

Even in the presence of this additional contribution to the total width, all modes in the range of validity of the theory ($m_n \lesssim M_5$) are narrow $\Gamma_n \ll m_n$ (see eq.~\eqref{approx-KKG-KKG}). Moreover, as long as 
\be
k \gtrsim 1.5\times 10^{-6}\, M_5 \approx 15~\mbox{MeV}\left(\frac{M_5}{10~\mbox{TeV}}\right) ,
\ee
the width is also smaller than the mass splitting between modes, allowing us to treat the KK excitations as individual resonances.
Finally notice that the additional contribution to the decay rate coming from KK self interactions does not preclude displaced decays, since only modes corresponding to large $n$ are affected.
\section{Results and conclusion}
\begin{figure}[t]
\begin{center}
\vspace{6mm}
\includegraphics[width=0.39\textwidth]{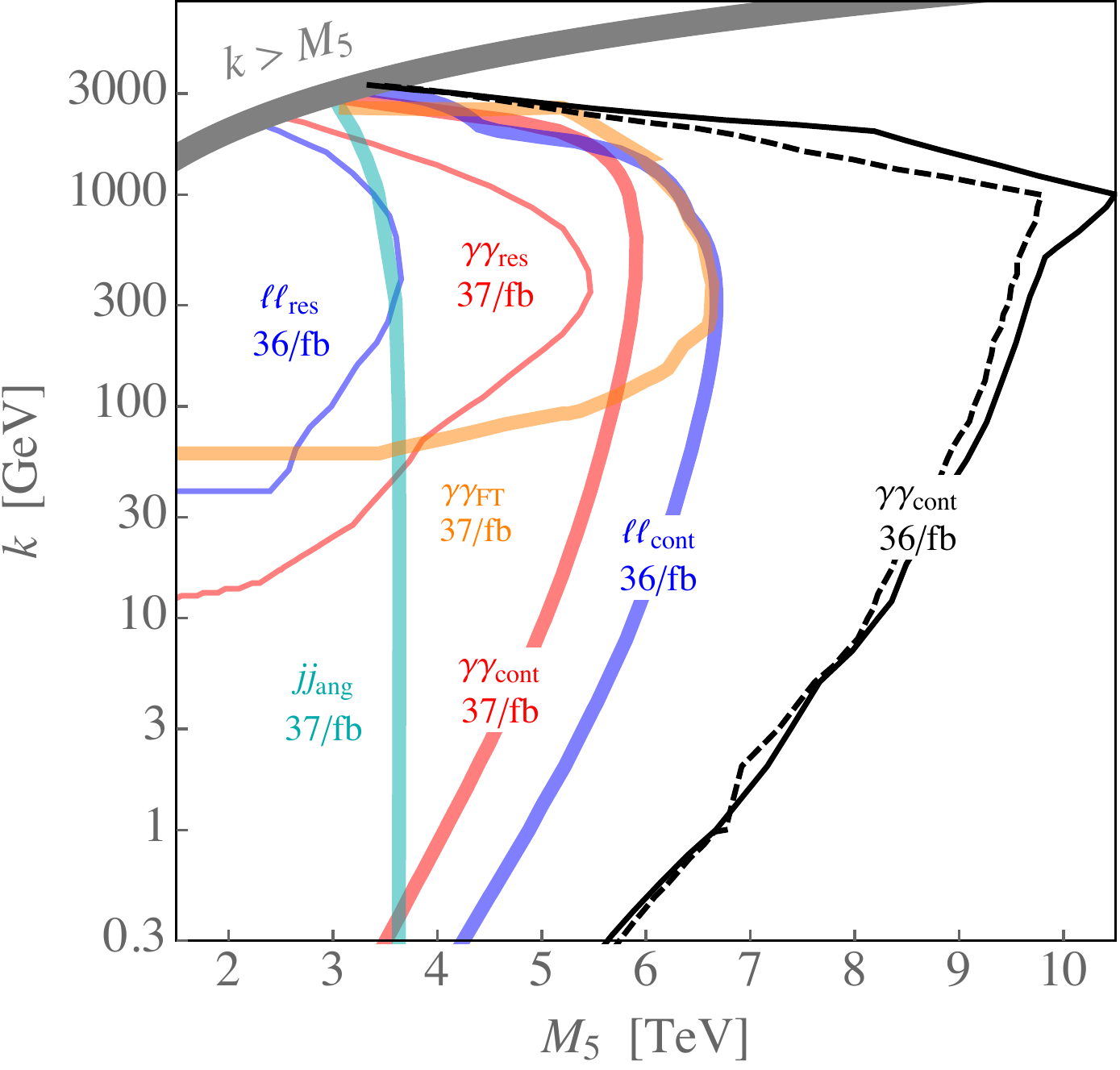}\\
\vspace{-4mm}
\end{center}
\caption{Summary of the constraints coming from the searches listed in the text. The black curves refer to the new CMS analysis \cite{CMS:2018thv}, with the dashed and solid ones being the expected and observed limits, respectively.}
\label{sensitivity}
\end{figure}

We have considered the CW/LD geometry and studied, among others, constraints from the following existing and proposed searches:
\begin{itemize}
\item Continuum $s$-channel effects in the invariant mass spectrum of the di-lepton~\cite{Aaboud:2017buh} and di-photon~\cite{Aaboud:2017yyg} channels;
\item Continuum $t$-channel effects in the di-jet angular distribution~\cite{Aaboud:2017yvp};
\item Resonant searches in the di-lepton~\cite{Aaboud:2017buh} and di-photon~\cite{Aaboud:2017yyg,CMS:2017yta} channels;
\item Fourier analysis of the di-photon invariant mass spectrum (for more details on this newly proposed technique see ref.~\cite{Giudice:2017fmj});
\item The new di-photon invariant mass spectrum analysis by CMS \cite{CMS:2018thv} appeared after ref.~\cite{Giudice:2017fmj}.
\end{itemize}
Our estimates of the sensitivity of these searches is summarized in figure~\ref{sensitivity}. Other constraints from $e^+e^-$ collisions at LEP, beam dump experiments, astrophysics, and BBN are not relevant in the region of the parameter space that we consider \cite{Giudice:2017fmj}. Finally, notice how the new CMS analysis of the continuum di-photon spectrum  \cite{CMS:2018thv}, whose constraint is represented by the rightmost black line in figure~\ref{sensitivity}, has already impressively overcome all other constraints, starting to test a more fine-tuned region of the parameter space.

\vspace{-3mm}
\section*{Acknowledgments}
\vspace{-3mm}
I thank Gian F. Giudice, Yevgeny Kats, Matthew McCullough, and Alfredo Urbano for collaboration on this project. I also thank Brando Bellazzini, Duccio Pappadopulo, Riccardo Rattazzi, Javi Serra, and Luca Vecchi for useful discussions on extra-dimensional models.
%

\vspace{-2mm}
\section*{References}

\bibliographystyle{mine}
\bibliography{torre_riccardo}
%
%
%
%

\end{document}